\documentclass[structabstract]{aa}  

\usepackage{graphicx}
\usepackage{txfonts}
\begin{document}
\title{Revision of the $^{15}$N(p,$\gamma$)$^{16}$O reaction rate and oxygen abundance in 
H--burning zones}
\author{A.~Caciolli
\inst{1,}
\inst{2}
\and
C.~Mazzocchi\thanks{present address: Inst. of Experimental Physics
University of Warsaw
ul. Hoza 69
00--682 Warszawa, Poland}
\inst{3}
\and
V.~Capogrosso
\inst{3}
\and
D.~Bemmerer
\inst{4}
\and
C.~Broggini
\inst{1}
\and
P.~Corvisiero
\inst{5}
\and
H.~Costantini
\inst{5}
\and
Z.~Elekes
\inst{6}
\and
A.~Formicola
\inst{7}
\and
Zs.~F\"ul\"op
\inst{6}
\and
G.~Gervino
\inst{8}
\and
A.~Guglielmetti
\inst{3}
\and
C.~Gustavino
\inst{7}
\and
Gy.~Gy\"urky
\inst{6}
\and
G.~Imbriani
\inst{9}
\and
M.~Junker
\inst{7}
\and
A.~Lemut\thanks{present address: Nuclear Science Division,
Lawrence Berkeley National Laboratory, 
1 Cyclotron Rd MS 88--R0192, 
Berkeley, CA 94720--8101, USA}
\inst{5}
\and
M.~Marta\thanks{present address: GSI Helmholtzzentrum f\"ur Schwerionenforschung, 64291 Darmstadt, Germany}
\inst{4}
\and
R.~Menegazzo
\inst{1}
\and
S.~Palmerini
\inst{10}
\and
P.~Prati
\inst{5}
\and
V.~Roca
\inst{9}
\and
C.~Rolfs
\inst{11}
\and
C.~Rossi~Alvarez
\inst{1}
\and
E.~ Somorjai
\inst{6}
\and
O.~Straniero
\inst{12}
\and
F.~Strieder
\inst{11}
\and
F.~Terrasi
\inst{13}
\and
H.~P.~Trautvetter
\inst{11}
\and
A.~Vomiero
\inst{14}
}

 \institute{ Istituto Nazionale di Fisica Nucleare (INFN), Sezione di Padova, via Marzolo 8, I--35131 Padova, Italy
 \and
 Dipartimento di Scienze della Terra, Universit\`a di Siena, I--53100 Siena, and Centro di GeoTecnologie CGT, I--52027 San Giovanni Valdarno, Italy
 \and
 Universit\`a degli Studi di Milano and INFN, Sezione di Milano, I--20133 Milano, Italy
 \and
 Helmholtz--Zentrum Dresden--Rossendorf, Bautzner Landstr. 400, 01328 Dresden, Germany
 \and
 Universit\`a di Genova and INFN Sezione di Genova, Genova, I--16146 Genova, Italy
 \and
 Institute of Nuclear Research (ATOMKI), H--4026 Debrecen, Hungary
 \and
 INFN, Laboratori Nazionali del Gran Sasso (LNGS), I--67010 Assergi (AQ), Italy
 \and
 Dipartimento di Fisica Sperimentale, Universit\`a di Torino and INFN Sezione di Torino, I--10125 Torino, Italy
 \and
Dipartimento di Scienze Fisiche, Universit\`a di Napoli ÒFederico II,Ó and INFN
Sezione di Napoli, I--80126 Napoli, Italy 
\and
Dipartimento di Fisica, Universit\`a  degli studi di Perugia and INFN, Sezione di
Perugia, I--06123, Perugia Italy
\and
Institut f\"ur Experimentalphysik, Ruhr--Universit\"at Bochum, D--44780 Bochum,
Germany
\and
INAF--Osservatorio Astronomico di Collurania, I--64100 Teramo, Italy
\and
Seconda Universit\`a di Napoli, I--81100 Caserta, and INFN Sezione di Napoli, I--80126
Napoli, Italy
\and
CNR IDASC SENSOR Lab and Dipartimento di Chimica e Fisica per l'Ingegneria e per i Materiali, Universit\`a di Brescia, Brescia, Italy
}

  \abstract
   { The NO cycle takes place in the deepest layer of a H--burning core or shell, when the temperature exceeds $T \simeq 30 \cdot 10^6$ K.
   The O depletion
observed in some globular cluster giant stars,   always associated with a Na enhancement, may be due to either a deep mixing  during the RGB (red giant branch) phase of the star or to the pollution of the primordial gas by an early population of massive AGB (asymptotic giant branch) stars, whose chemical composition was modified by the hot bottom burning. In both cases, the NO cycle is responsible for the O depletion. }
{The activation of this cycle depends on the rate of the $^{15}$N(p,$\gamma$)$^{16}$O reaction. A precise evaluation of this reaction rate at temperatures as low as experienced in H--burning zones in stellar interiors is mandatory to understand the observed O abundances.}
{We present a new measurement of the $^{15}$N(p,$\gamma$)$^{16}$O reaction performed at LUNA covering for the first time the center of mass energy range 70--370 keV, which  corresponds to stellar temperatures between 65 $\cdot 10^6$ K and 780 $\cdot 10^6$ K. This range includes the $^{15}$N(p,$\gamma$)$^{16}$O Gamow--peak energy of explosive H--burning taking place in the external layer of a nova and the one of the hot bottom burning (HBB) nucleosynthesis occurring in massive AGB stars.}
{With the present data,  we are also able to confirm the result of the previous R--matrix extrapolation. In particular, in the temperature range of astrophysical interest, the new rate is about a factor of 2 smaller than reported in the widely adopted compilation of reaction rates (NACRE or CF88)  and the uncertainty is now reduced down to the 10\% level.
}
   {}

   \maketitle
%
\section{Introduction}
Hydrogen burning in stars proceeds through two different sets of nuclear reactions: the
proton proton (pp) chain and the carbon nitrogen oxygen (CNO) cycle. 
While in low mass main sequence stars the
energy supply is provided by the pp--chain\footnote{The pp--chain also dominates the H burning 
in extremely--metal--poor stars of any mass, due to the 
lack of C, N and O nuclei}, the CNO cycle is the principal nuclear process in the core of high mass
main sequence stars ($M$~$\gtrsim $~1.2~$M_\odot$) as well as in the H--burning shell of giant stars
\cite{iben67}. Furthermore, a hot CNO cycle may occur at the surface of H--accreting compact objects, 
like white dwarfs or neutron stars \cite{jose1998}. 
\begin{figure}[htbp]
\begin{center}
\includegraphics[scale=0.30]{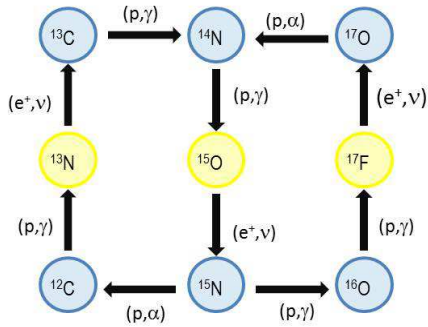}
\caption{The CNO bi--cycle.}
\label{cno}
\end{center}
\end{figure}

The set of nuclear reactions involved in the CNO cycle is illustrated in Fig. \ref{cno}. 
Actually, it is a combination of two distinct cycles,
called CN and NO, respectively. The proton capture on $^{15}$N results in two possible
channels, the $^{15}$N(p,$\alpha$)$^{12}$C and the
$^{15}$N(p,$\gamma$)$^{16}$O, respectively: the ratio of  
the rates  provides the linkage between the CN and the NO cycles.
The CN cycle becomes fully active when the temperature attains 
$T_9 \gtrsim 0.016$--$0.020$\footnote{$T_9=T(K)/10^9$.},
while the NO cycle requires higher temperatures ($T_9 \gtrsim 0.030$--$0.035$\footnote{ 
The activation temperatures of both the CN and the NO cycles depend on 
the actual amount of C, N and O nuclei and, therefore, on the stellar metallicity.}). 
In case of an active NO cycle, this process determines the abundances of all
the stable oxygen isotopes ($^{16}$O, $^{17}$O, $^{18}$O).
For this reason, a precise evaluation of the $^{15}$N(p,$\gamma$)$^{16}$O reaction rate is needed to
address several astrophysical problems,
like deep mixing scenarios in red giant stars \cite[see e.g.][]{sweigart1979, langer1986, charb98, kraft93,
boo95, wass95, denis2003, pal10}, hot bottom burning nucleosynthesis in massive AGB stars \cite{renzini1981} or the
H--burning nucleosynthesis in nova--like events \cite{Iliadis2002,Jose2007}.

At low energies the cross section $\sigma (E)$ of the $^{15}$N(p,$\gamma$)$^{16}$O  reaction ($Q$--value = 12.127 MeV) is
typically expressed in terms of the astrophysical S--factor $S(E)$ defined for this reaction as:
\begin{equation}\label{eq:Sfactor}
S(E) = \sigma (E) E \exp(212.85 / \sqrt{E})
\end{equation}
where $E$ is the center of mass energy in keV.

In hydrostatic H--burning, the Gamow peak energy of this reaction ranges between
30 and 100 keV. Larger values, up to 300 keV, may be attained during explosive burning.
In this energy range, the astrophysical S--factor is
influenced by two resonances at $E$ = 312 and 964 keV\footnote{In the center of mass reference. Beam energies are given in the center of mass reference unless otherwise stated} related to excited states in $^{16}$O at $E_x$ = 12440 and
13090 keV, respectively.
The reaction rates reported in the NACRE \cite{NACRE} and the CF88 \cite{Caughlan1988} compilations 
are  based on the direct measurement presented by Rolfs and Rodney (1974).
However, more recent R--matrix studies  \cite{Mukhamedzhanov2008,Barker2009}, which also take  into account a previous ANC measurement \cite{Mukhamedzhanov2008}, suggested a substantial reduction of the S(0) (i.e. the astrophysical factor at $E=0$). 
This result is in agreement with older direct measurements \cite{Hebbard1960, Brochard1973}. 

This discrepancy prompted an in--depth study of the reaction at LUNA 
(Laboratory for Underground Nuclear Astrophysics).
The LUNA facility has been designed to study nuclear reactions of astrophysical interest
at the same energies of the stellar interiors, 
by taking advantage of the ultra--low background \cite{Bemmerer2005,Caciolli2009} of the INFN--Gran Sasso underground laboratory 
(a detailed description of LUNA and its experimental study of the 
pp chain and CNO cycle may be found in the following reviews: \cite{costantini2009, broggini2010}). 
First of all, a re--analysis of data taken 
with nitrogen gas target of natural isotopic composition (0.4\% $^{15}$N) at  $E$ = 90--230 keV has been
performed \cite{Bemmerer2009}. Then, a new measurement has been carried out
at LUNA and Notre Dame \cite{LeBlanc2010}. HPGe detectors and enriched TiN solid targets have been
used to cover a wide energy range, namely: $E$ = 120--1800 keV.
Although the minimum energy is still too high to study most of the stellar H--burning environments,
thanks to the excellent accuracy (7\%) and the wide energy range,
this new experiment provided a dataset suitable for an R--matrix extrapolation toward lower energies.

In this paper, we present a third experiment performed at LUNA,
designed to explore lower energies.
The use a BGO detector, having a higher $\gamma$--detection efficiency compared to the HPGe detectors,
allowed us to easily cover the 312 keV resonance region and to extend the direct measurements
down to 70 keV. The aim of this further effort is twofold. First of all, the new data set
covers the Gamow peak corresponding to the explosive burning in Novae as well as 
hot bottom burning in massive AGB stars. 
Furthermore, it provides an independent test of the low energy R--matrix extrapolation.

In the next section we illustrate the  experiment, the data analysis and the results.
In particular, a comparison of the present, independent measurement with the
low energy predictions of the R--matrix analysis \cite{LeBlanc2010},
leads to the conclusion that the  $^{15}$N(p,$\gamma$)$^{16}$O reaction rate is now known within a 10\%
confidence interval.
A summary of the astrophysical studies requiring an accurate evaluation of 
the $^{15}$N(p,$\gamma$)$^{16}$O reaction rate follows.

\section{The new underground experiment}\label{sec2}

The target and the $\gamma$--ray detection set--up are those  used in previous measurements and have been already extensively described elsewhere (for instance see \cite{Limata2010}). The proton beam (30--150 $\mu$A) reaches the water cooled target after passing a 5 mm diameter collimator and a 1 m long copper tube, which is cooled  to liquid nitrogen temperatures and works as a cold trap in order to prevent  impurities scattered by the beam from depositing on the target surface. The pressure in the target chamber is $5 \cdot 10^{-7}$ mbar and no carbon deposition on the target   is observed after the irradiation. This is checked  by performing scans of the profile of the $^{14}$N(p,$\gamma$)$^{15}$O resonance at $E_p$ = 278 keV. The target chamber works as a Faraday cup and provides the  integral of the charge deposited, hence the average beam intensity, with an overall uncertainty of 2\% (a $-$300 V high voltage is applied to the cold trap to suppress the secondary electron emission).
\begin{figure}[htbp]
\begin{center}
\includegraphics[angle=0,width=0.45\textwidth]{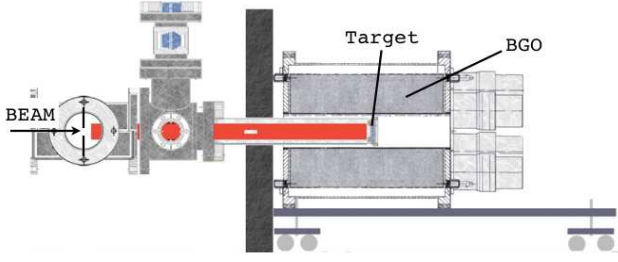}
\caption{Schematic representation of the last portion of the beam--line with the detection set--up.}
\label{fig:set-up}
\end{center}
\end{figure}

The target is surrounded by a 4$\pi$--BGO summing crystal (28 cm long, 20 cm diameter, and 6 cm coaxial hole, \cite{Casella2002}). The 4$\pi$--BGO is essential in order to increase the $\gamma$--detection efficiency, 
which is calculated with a simulation based on GEANT4 \cite{Agostinelli2003} and carefully checked 
with radioactive sources and with the $\gamma$--ray produced by the proton induced 
reaction $^{11}$B(p,$\gamma$)$^{12}$C at the $E$ = 149 keV resonance.  
The simulation needs experimental inputs, such as the decay scheme and the angular distribution 
of the emitted $\gamma$--radiation. 
The decay branching ratios for transitions to the excited state of 
$^{16}$O have been measured by Rolfs and Rodney (1974), 
Bemmerer et al. (2009) and LeBlanc et al. (2010). 
The angular distribution has been found to be isotropic in a previous LUNA work  \cite{LeBlanc2010}.
By considering all the contributions described above in the simulation code, the total uncertainty  on the efficiency is  3\%.

The TiN forming the target material, enriched in $^{15}$N, is deposited on a tantalum backing with the reactive sputtering technique \cite{Rigato2001}. The target thickness is 100 nm, as verified  through secondary neutral mass spectroscopy \cite{SIMS} (the uncertainty on this measurement is included in the contribution to the target analysis in Table \ref{tab:errorSolid}),
corresponding to 15 keV energy loss at $E$ =  259 keV. The stoichiometry Ti/N, which  ranges from 
0.97 to 1.18 according to the target, is measured for each target with the high Z elastic recoil 
detection (ERD) technique \cite{Bergmaier1998}. Isotopic abundances between 96\% and 99\%,  
according to the target, are deduced from the observed height of the plateau in the yield of the
$^{14}$N(p,$\gamma$)$^{15}$O resonance at $E$ = 259 keV, and from the ERD data. 
The results from these two methods agree within 2\%. Finally, the target deterioration, 
caused by the impinging high--intensity proton beam  has been studied by using the 
430 keV resonance of $^{15}$N(p,$\alpha \gamma$)$^{12}$C \cite{Marta2010}. 
The targets have been analyzed  by looking at the shape of the plateau in the yield distribution 
for the 430 keV resonance. The surface irradiated by the LUNA beam and the area outside the LUNA 
beam--spot are investigated, so that appropriate corrections for the target deterioration 
during measurements are derived.

\begin{figure}[htbp]
\centering
\includegraphics[angle=0,width=0.5\textwidth]{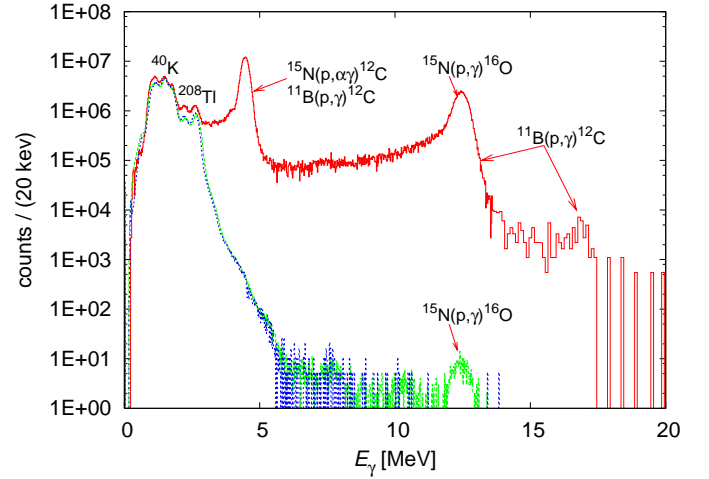}
\caption{Spectra taken  at $E_\mathrm{p}$ = 330 keV (red), $E_\mathrm{p}$ = 80 keV  (green), and laboratory
background spectrum (blue), normalized to the same time.}
\label{fig:spectrum}
\end{figure}

The laboratory background in the $^{15}$N(p,$\gamma$)$^{16}$O region of interest   is  about 6~counts/day. The beam induced background in the same region, produced by the $^{11}$B(p,$\gamma$)$^{12}$C reaction, is monitored  by means of the peak produced by this reaction at 16.1 MeV (see Fig. \ref{fig:spectrum} as an example of the acquired spectra). The counts in this peak are usually  more than in the 11~MeV peak produced by the same reaction, which  lies within the $^{15}$N(p,$\gamma$)$^{16}$O region of interest of the spectrum \cite{Bemmerer2009}. We have rejected all measurements where the 16 MeV peak  contained more than 3\% of the counts in our region of interest. \\
\begin{table}
\caption{The absolute S--factor data and their statistical uncertainties from the present work. The systematic error is 11\%.}
\label{tab:Sfactor}
\centering
\begin{tabular}[htb]{ccc|ccc}
\hline \hline
 $E$ 	&	$S$  & $\Delta S^\mathrm{stat}$ &  $E$		&	$S$  & $\Delta S^\mathrm{stat}$ \\
  $[$keV] & [keV barn] & [keV barn] & [keV] & [keV barn] & [keV barn] \\
 \hline
72.8  & 52 	  & 4	& 236.9 &	153.0 & 1.1	\\
81.3  & 49 	  & 2	& 246.3 &	172.0 & 1.7	\\
89.3  & 53 	  & 6	& 256.4 &	201   & 3	\\
105.1 &	59 	  & 4	& 266.3 &	227   & 3	\\
114.8 &	53 	  & 3	& 274.5 &	254.7 & 1.4	\\
123.5 &	56.4  & 1.8 &	283.5 &	283   & 3	\\
132.7 &	64 	  & 2	& 293.7 &	315.4 & 1.6	\\
143.7 &	68.3  & 1.0 & 	302.6 &	320   & 2	\\
151.3 &	55.9  & 1.1 &	311.7 &	309.1 & 1.2	\\
162.3 &	79.2  & 0.5 &	321.1 &	277.6 & 1.0	\\
170.7 &	79.8  & 0.9 &	330.4 &	227.4 & 0.7	\\
180.1 &	87.2  & 1.0 &	340.2 &	183.0 & 0.9	\\
189.1 &	93.5  & 1.1  & 349.1 &	134.0 & 0.9	\\
198.4 &	102.6 & 0.8 &	354.1 &	124.0 & 1.2	\\
207.9 &	114.0 & 1.6 &	358.8 &	101.0 & 1.0	\\
217.3 &	123.0 & 1.0 &	363.6 &	95.1  & 0.9	\\
227.4 &	136.0 & 1.1 &	368.3 &	81.0  & 0.8	\\
\hline \hline
\end{tabular}
\end{table}
The target profiles can be integrated with the cross section in order to calculate the expected
yield as:
\begin{equation}\label{eq:solidY}
Y_{sim} =  \int^{x_\mathrm{max}}_{x_0} S(E_\mathrm{p}) \cdot E_\mathrm{p} \exp \left(\frac{212.85}{\sqrt{E_\mathrm{p}}}\right)   \eta_\mathrm{BGO}
\cdot n_\mathrm{target} (x) \frac{^{15}\mathrm{N}}{\mathrm{N}} \mbox{d}x
\end{equation}
where $E_\mathrm{p}$ is the energy in the laboratory system expressed in keV and it depends on the beam position $x$ along the target thickness, $\eta_\mathrm{BGO}$ is the efficiency, and $n_\mathrm{target} (x) \frac{^{15}\mathrm{N}}{\mathrm{N}}$ is the number of $^{15}$N nuclides in the $x$ position in the target.
By comparing the experimental yield $Y_\mathrm{exp}$ with the calculated one, it is possible to determine
the S--factor as follows:
\begin{equation}\label{eq:solidSfactor}
S(E_\mathrm{eff})_\mathrm{exp} = \frac{Y_\mathrm{exp}}{Y_\mathrm{sim}} \cdot S(E_\mathrm{eff})_\mathrm{th}
\end{equation}
where the effective energy is calculated according to the following definition
\cite{Lemut2008}:
\begin{equation}\label{eq:effE}
E_\mathrm{eff} = \frac{\int^{x_\mathrm{max}}_{x_0} S(E) \cdot E \exp\left(\frac{212.85}{\sqrt{E}}\right) \cdot n_\mathrm{target} (x)
\cdot  \frac{^{15}\mathrm{N}}{\mathrm{N}} \cdot E \cdot \mbox{d}x}{\int^{x_\mathrm{max}}_{x_0} S(E) \cdot E
\exp\left(\frac{212.85}{\sqrt{E}}\right)  \cdot n_\mathrm{target} (x) \cdot  \frac{^{15}\mathrm{N}}{\mathrm{N}} \mbox{d}x}.
\end{equation}
In Eq. (\ref{eq:effE}) the theoretical S--factor  is used. Four different theoretical S--factors are considered in Eq. (\ref{eq:solidY}) and Eq. (\ref{eq:effE}): the one reported in \cite{LeBlanc2010} and the one reported in \cite{Mukhamedzhanov2011}, a constant S--factor and a value obtained from a recursive analysis process. In all cases,  the same results are obtained within 1\% discrepancies which is included in the error on the effective energy.

As reported in Table \ref{tab:Sfactor}, the $^{15}$N(p,$\gamma$)$^{16}$O astrophysical S--factor 
is obtained for the center of mass energy range 70--370 KeV.
The statistical uncertainty is always limited within a few percent, 
reaching a maximum value of 10\% at $E$ = 72.8 keV. 
All sources of systematic uncertainties are given in Table \ref{tab:errorSolid} and sum to a total systematic uncertainty of 10\%.
\begin{table}
\centering
\caption{S--factor systematic uncertainties.}
\label{tab:errorSolid}
\begin{tabular}[!htb]{lc}
\hline \hline
Source      & Estimated     \\
description & uncertainty	 \\
 \hline
Target analysis 			&	7.5\%	\\
stopping power & 4.0\% \\
$^{15}$N isotopic ratio	&	2.0\%	\\
Ti/N stoichiometry		&	2.0\%	\\
Beam intensity			 &	2.0\%	\\
Effective energy		&	3.0\%	\\
$\gamma$--ray detection efficiency	&	3.0 \%	\\
$^{11}$B(p,$\gamma$)$^{12}$C background &	3.0\%	\\
Total systematic uncertainty	&	10.0\%	\\
\hline \hline
\end{tabular}
\end{table}

\begin{figure}[htbp]
\centering
\includegraphics[angle=0,width=0.5\textwidth]{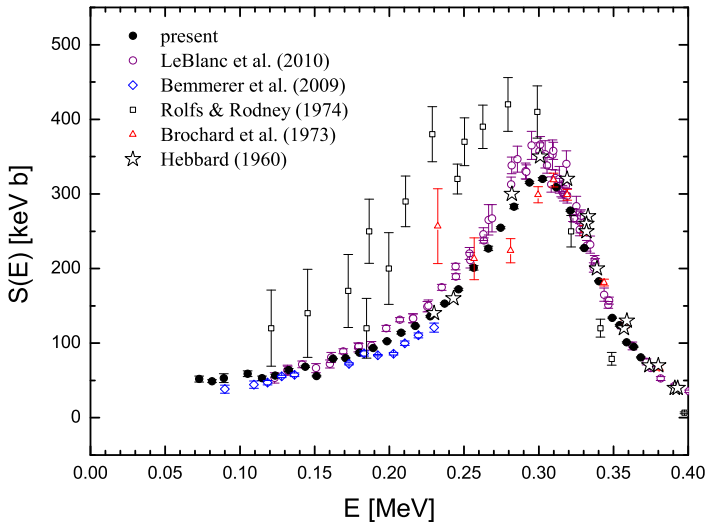}
\caption{The S--factor as a function of energy. Present data (black dots) are compared to the results 
of previous experiments.}
\label{fig:Sfactor}
\end{figure}

\begin{figure}[htbp]
\centering
\includegraphics[angle=0,width=0.45\textwidth]{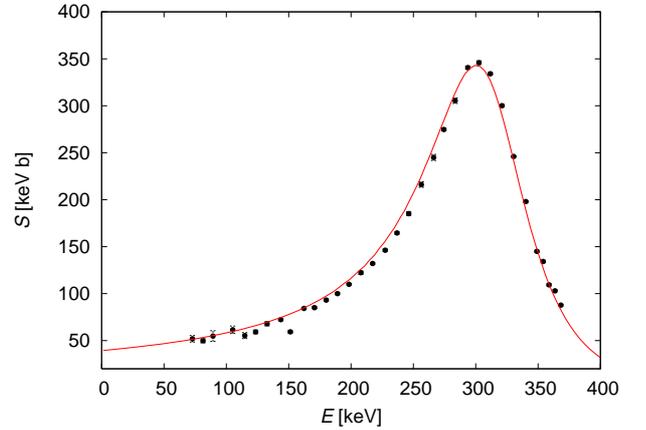}
\caption{The present data rescaled to the new recommended value of $\sigma(312)$ keV (see text to details) and compared to the R--matrix predictions \cite{LeBlanc2010}.}
\label{fig:rescale}
\end{figure}

A comparison of the derived astrophysical S--factor to the results of previous experiments is
shown if Fig. \ref{fig:Sfactor}. 
We confirm the previous finding concerning the need of a substantial reduction of the S(0) value.
The present result is significantly lower than the resonant cross section from Rolfs and Rodney (1974), 
i.e. the data set adopted in NACRE and CF88, and, by considering the systematic uncertainty, 
in good agreement with our previous HPGe measurement \cite{LeBlanc2010}.  
In particular, according to the present absolute analysis,
 the cross section on top of the $E=312$~keV resonance 
 is $\sigma(312~keV)=6.0\pm0.6~\mu$b, where 
the quoted error includes the 10~\% systematic uncertainty. In Table \ref{table:res_cross}, 
we compare this result to the values of previous measurements. 
The weighted average of 3 measurements\footnote{The result obtained by \cite{Brochard1973} 
has been excluded, because no uncertainty was reported.} 
leads to a recommended value of $\bar\sigma(312~keV)=6.5\pm0.3~\mu$b. 
The shape of the R--Matrix fit has been also compared to the present data as shown in Fig. \ref{fig:rescale}. Only for this comparison the present data have been corrected for the electron screening in the adiabatic approximation \cite{Assenbaum1987} (at most 10\% at 70 keV) and they have been  rescaled to the calculated average value. This rescaling is still between inside the systematic uncertainties of the present absolute data.
They show an excellent agreement with the energy dependence of the LUNA R--Matrix fit \cite{LeBlanc2010}.

\begin{table}[tbh]
\caption{\label{table:res_cross} Summary of 312~keV resonance cross sections in
comparison to previous results (see text for details and  references). The uncertainty reported by Hebbard (1960) has been obtained by assuming it to be 10\% as reported by \cite{Barker2009}.}
\centering
\begin{tabular}{ccccc}
\hline \hline
Present & LeBlanc      & Rolfs and & Brochard & Hebbard \\
study   & {\it et al.} & Rodney    & {\it et al.} & \\
 
 [$\mu$b] & [$\mu$b] & [$\mu$b] & [$\mu$b] & [$\mu$b] \\
\hline
$6.0\pm0.6$ & $6.5\pm0.3$ & $9.6\pm1.3$ & 6.3 & $6.5\pm0.7$ \\
\hline \hline

\end{tabular}
\end{table}

Finally, a new R--matrix 
analysis has been recently published by Mukhamedzhanov, La Cognata and Kroha (2011).
By varying the fitting method, these authors obtain $S(0)$ values ranging 
between  33.1 and 40.1 keVb, which is in excellent agreement with the value reported 
by LeBlanc et al. ($S(0)=39.6 \pm 2.6$ keVb).  

For practical purposes, the nuclear reaction rate can be approximated by the following 
fitting formula \cite{LeBlanc2011}:

\begingroup
\begin{eqnarray}
N_A <\sigma v> &=& a_110^9T^{-\frac{2}{3}}\exp[a_2 T^{-\frac{1}{3}}-(T/a_3)^2] \\ \nonumber
 & & [1 + a_4T + a_5T^2] + a_610^3 T^{-\frac{3}{2}}\\\nonumber
 & & \exp(a_7/T) + a_810^6T^{-\frac{3}{2}}\exp(a_9/T),
\end{eqnarray}
\endgroup
where the best fit parameters are reported in Table~\ref{table:rr_fit_par}.

\begin{table}[tbh]
\caption{\label{table:rr_fit_par} Best fit parameters for the
$^{15}$N(p,$\gamma$)$^{16}$O reaction rate given in \cite{LeBlanc2010}.}
\centering
\begin{tabular}{lll}
\hline \hline
a$_1$ =   0.523  & a$_4$ = 6.339  & a$_7$ = $-$2.913  \\
a$_2$ = $-$15.240  & a$_5$ = $-$2.164 & a$_8$ =  3.048 \\
a$_3$ =   0.866  & a$_6$ = 0.738  & a$_9$ = $-$9.884 \\
\hline \hline
\end{tabular}

\end{table}

\section{Summary and Conclusions}
\begin{figure}[htbp]
\centering
\includegraphics[angle=0,width=0.45\textwidth]{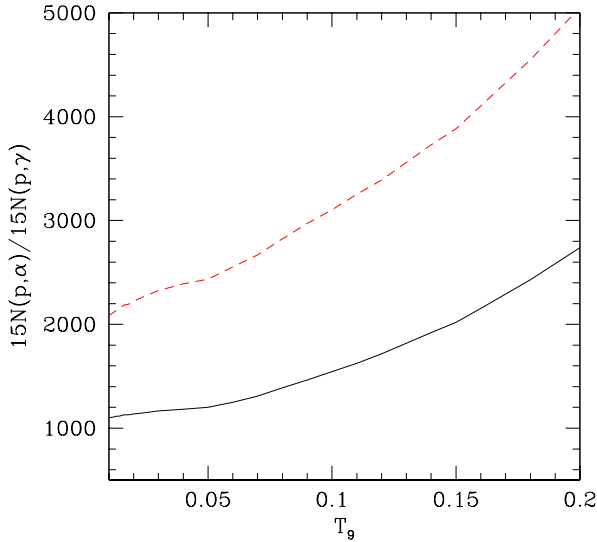}
\caption{The CN--NO branching ratio, as a function of the temperature, under different 
assumptions for the $^{15}$N(p,$\gamma$)$^{16}$O reaction rate: 
NACRE (solid line) and the revised rate (dashed line).}
\label{branch}
\end{figure}

\begin{figure}[htbp]
\centering
\includegraphics[angle=0,width=0.45\textwidth]{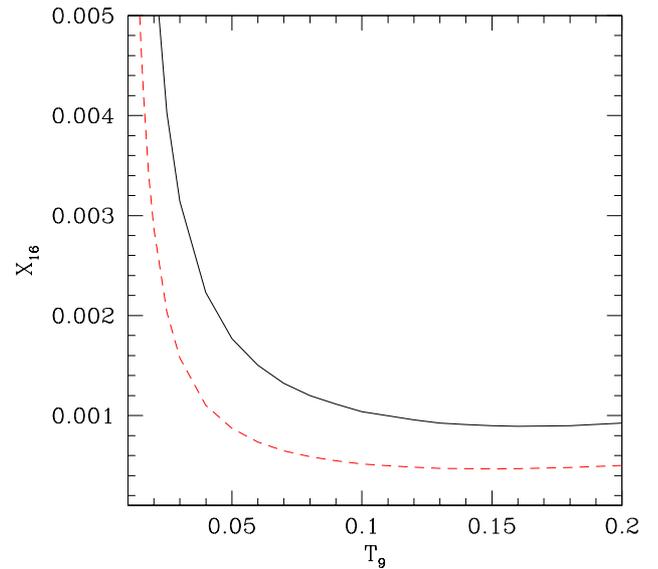}
\caption{The $^{16}$O equilibrium abundance (mass fraction) as given by:
$X_{16} = X_{15} {16 \over 15}{<\sigma v>_{15N+p} \over <\sigma v>_{16O+p}}$.
The solid and the dashed lines represents the old (NACRE) and the
new (revised) prescriptions for the $^{15}$N(p,$\gamma$)$^{16}$O reaction rate. A solar $^{15}$N
mass fraction ($X_{15}$), as derived by
Asplund et al. 2010, has been used. In both cases, NACRE prescriptions
for the $<\sigma v>_{16O+p}$ have been adopted.}
\label{equi}
\end{figure}

In this paper we have discussed the experimental efforts done to improve 
our knowledge of the $^{15}$N(p,$\gamma$)$^{16}$O reaction rate in the temperature range experienced by any H--burning zone in stellar interiors.
Such an important reaction is located at the branching point between the CN and NO cycles. The branching ratio, as a function of the temperature, is shown in Fig. \ref{branch}, where the solid line has been obtained by means of the widely adopted reaction rate  given by NACRE, while the dashed line represents the revised scenario as derived from the latest R--matrix study (see section \ref{sec2}).
In both cases, the rate suggested by NACRE has been used for the competitive 
$^{15}$N(p,$\alpha$)$^{12}$C reaction\footnote{This reaction has been recently studied  with the THM method \cite{cognata2009}. The authors do not report a reaction rate but only the $S(0)$ value. Scaling the previous NACRE results on that value the following considerations do not change so we still adopt the NACRE results in the present work}.
A look at the solid line shows that in the whole range of temperatures experienced
by the core and the shell--H burning, the $\alpha$--channel is between 1000 to 2000 times more
efficient than
the $\gamma$ channel: just 1 to 2 protons out of every 2000 are consumed by the NO cycle.
When the updated rate for the
$^{15}$N(p,$\gamma$)$^{16}$O is adopted, such a ratio becomes about a factor of 2 larger.
Although such a variation has negligible consequences on the overall nuclear energy production,
a change in the rate of the  $^{15}$N(p,$\gamma$)$^{16}$O affects the equilibrium abundances
of the stable oxygen isotopes within the H burning zone.
As an example, the equilibrium abundance of $^{16}$O is reported as
a function of the temperature in Fig. \ref{equi}.
Also in this case, the solid and the dashed lines represent the values obtained by adopting the
NACRE and the revised rate of the $^{15}$N(p,$\gamma$)$^{16}$O reaction, respectively.

Let us point out that the most important improvement resulting from the present analysis
of the CN--NO branching concerns the significant reduction of the nuclear physics uncertainties,
other that the change of the reaction rate with respect to the values
reported by CF88 or NACRE. For
stellar models and nucleosynthesis calculations implying H--burning whose Gamow peak energy is
larger than the minimum value attained by the LUNA BGO experiment, namely $E_0>70$ KeV, which
corresponds to a temperature $T>65 \cdot 10^6$ K, a true experimental error (smaller than 10\%) is now
available for this important reaction rate. 
Note that only in a very few cases the reaction rate has been measured
down to the stellar Gamow peak energy (see, e.g., \cite{Bonetti1999}). 
In  addition, basing on the good agreement found between the new LUNA measurements 
and the revised R--matrix fit (see previous section), we are confident that
the quoted small uncertainty may be assumed also in the extrapolated region.

Among the many astrophysical applications of the present analysis, we recall the explosive
H--burning in Novae, which occurs at temperature larger than $10^8$ K and, therefore, well above
the achieved experimental limit. A recent study by Iliadis et al. (2002), investigates
the dependence of the nova nucleosynthesis calculations on the various nuclear physics inputs.
They found that a reduction of a factor of two of the  $^{15}$N(p,$\gamma$)$^{16}$O reaction rate
would imply a 30\% reduction of  the final oxygen abundance.
Also the inner region of the convective envelope of massive AGB stars attains quite high
temperature, up to $T_9=0.08$--$0.09$ \cite{renzini1981, Forestini1997, 
Dantona1996, Straniero2000, Lattanzio2000}. The
resulting H burning, the so called hot bottom burning, coupled to the convective mixing,
gives rise to a very promising nucleosynthesis scenario, where all
the C, N and O isotopes are substantially affected. If the temperature is large enough ($80\times
10^6$ K), the Ne--Na and the Mg--Al cycles are also activated.  In this context, it has been
recently claimed that massive AGB stars  played a fundamental role during the  early
evolution of globular clusters \cite{Ventura2001}. According to this self--enrichment scenario, 
in between 50 to 100 Myr after the cluster formation,
the first generation of intermediate mass stars (5--7 M$_\odot$) reached the AGB.
Then, during this evolutionary phase, they underwent a substantial modification of the
envelope composition, as a consequence of the HBB and several dredge up episodes. 
Due to the huge AGB mass loss, fresh gas enriched in He, C, N and Na, but O
depleted, refilled the space occupied by the young Globular Cluster. If the star formation
process was still active at that epoch, some of the stars we observed today should show
the imprint of such a delayed chemical pollution by massive AGB. In particular, the O--Na
anti--correlation, as observed in Giant, sub--Giant and turn--off stars of several globular clusters 
\cite[e.g.][and reference therein]{kraft1997, Carretta2009}, may be the
consequence of this nucleosynthesis process. Such a conclusion follows from the evidence that
the temperature required for the activation of the NO cycle is similar to that
required for the activation of the Ne--Na cycle. Thus, when O is depleted at the bottom of 
the convective envelope, Na should be enhanced. 
For this reason, a precise determination of the $^{15}$N(p,$\gamma$)$^{16}$O is
one of the prerequisites to obtain a robust prediction of the O abundance and, in
turn, to check  the proposed self--pollution scenario for the observed O--Na anti--correlation.           

The R--matrix studies also allow to extrapolate
the precise experimental measurements of the $^{15}$N(p,$\gamma$)$^{16}$O reaction rate down to the
temperature range experienced  by the H--burning taking place in main sequence, RGB and
less--massive AGB stars. Also in these cases the uncertainty has been significantly reduced. Such
an occurrence may be immediately translated in more robust astrophysical predictions.

\section*{Acknowledgments}
We thank  A. Bergmaier (Universit\"at der Bundenswehr M\"unchen) and Javier Garc\'ia Lopez (CNA,
Centro Nacional de Aceleratores) of Seville for assistance with the isotopic abundance analysis.
Financial support by INFN and in part by the European Union (TARI RII3--CT--2004-- 506222, AIM 025646
and SPIRIT 227012), the Hungarian Scientific Research Fund (K68801), and DFG (BE 4100/2--1) is
gratefully acknowledged.

\end{document}